\documentclass[11pt,a4paper]{article}

\usepackage{fullpage}
\usepackage{amsmath,amssymb,dsfont}
\usepackage{listings,hyperref}
\usepackage{graphicx,subfigure,xcolor}
\usepackage{amssymb}

\usepackage{algorithmic,algorithm}

\floatname{algorithm}{Procedure}

\algsetup{indent=1em}

\newcommand{\MMA}{\emph{Mathematica}}

\newcommand{\eg}{{\emph{e.g.\/}}}
\newcommand{\ie}{{\emph{ie.\/}}}

\newcommand{\Id}{\mathds{1}}
\newcommand{\tr}{\mathrm{tr}}

\DeclareMathOperator{\res}{\mathbf{res}}

\newcommand{\C}{\ensuremath{\mathds{C}}}
\newcommand{\Cplx}{\ensuremath{\C}}

\newcommand{\ComplexToReal}{\operatorname{\mathcal{R}}}
\newcommand{\RealToComplex}{\operatorname{\mathcal{C}}}

\begin{document}

\title{States and channels in quantum mechanics without complex numbers\footnote{Presented at 21${}^{\mathrm{st}}$ International Conference on Applications of Computer Algebra 2015 (ACA2015), July 20-23, 2015, Kalamata, Greece.}}
\author{J.A. Miszczak\footnote{E-mail: miszczak@iitis.pl}
\\ Institute of Theoretical and Applied Informatics, Polish Academy of
Sciences\\ Baltycka 5, 44100 Gliwice, Poland \and Applied Logic, Philosophy and History of Science group, 
University of Cagliari\\Via~Is Mirrionis 1, 09123 Cagliari, Italy}

\date{0.15 (15/03/2016)}

\maketitle

\abstract{In the presented note we aim at exploring the possibility of
abandoning complex numbers in the representation of quantum states and
operations. We demonstrate a simplified version of quantum mechanics in which
the states are represented using real numbers only. The main advantage of this
approach is that the simulation of the $n$-dimensional quantum system requires
$n^2$ real numbers, in contrast to the standard case where $n^4$ real numbers
are required. The main disadvantage is the lack of hermicity in the
representation of quantum states. Using \MMA\ computer algebra system we develop
a set of functions for manipulating real-only quantum states. With the help of
this tool we study the properties of the introduced representation and the
induced representation of quantum channels.}

\section{Introduction}
Quantum information theory aims at harnessing the behavior of quantum mechanical
objects to store, transfer and process information. This behavior is, in many
cases, very different from the one we observe in the classical
world~\cite{miszczak12high-level}. Quantum algorithms and protocols take
advantage of the superposition of states and require the presence of entangled
states. Both phenomena arise from the rich structure of the space of quantum
states~\cite{BZ06}. Hence, to explore the capabilities of quantum information
processing, one needs to fully understand this space. Quantum mechanics provides
us also with much larger of allowed operations than in classical case space. It
can be used to manipulate quantum states. However, the exploration of the
space of quantum operations is fascinating, but cumbersome task.

Functional programming is frequently seen as an attractive alternative to the
traditional methods used in scientific computing, which are based mainly on the
imperative programming paradigm~\cite{hinsen09promises}. Among the features of
functional languages which make them suitable for the use in this area is the
easiness of execution of the functional code in the parallel environments.

During the last few years \MMA\ computing system has become very popular in the
area of quantum information theory and the foundations of quantum mechanics. The
main reason for this is its ability to merge the symbolic and numerical
capabilities~\cite{WL}, both of which are often necessary to understand the theoretical
and practical aspects of quantum systems~\cite{qdensity, qcwave,
gerdt09mathematica, miszczak13functional}.

In this paper we utilize the ability to merge symbolical and numerical
calculations offered by \MMA\ to investigate the properties of the variant of
quantum theory based of the representation of density matrices built using
real-numbers only. We start by introducing the said representation, including
the \MMA\ required functions. Next, we test the behavior of selected partial
operations in this representation and consider the general case of quantum
channels acting on the space of real-only density matrices. In the last part we
provide some insight into the spectral properties of the real-only density
matrices. Finally, we provide the summary and the concluding remarks.

\subsection{Preliminaries}\label{sec:preliminaries}
In quantum mechanics the state is represented by positive semidefinite,
normalized matrix. In the following we focus on this property as it is
crucial for the properties of quantum states and channels. To be more specific,
we aim at using symbolic matrix which are hermitian. Using the symbolic
capabilities of \MMA\ they can be expressed as
\begin{lstlisting}
SymbolicDensityMatrix[a_, b_, d_] := Array[
  If[#1 < #2, $a_{\#1, \#2}$ + I $b_{\#1, \#2}$, 
  	If[#1 > #2, $a_{\#2, \#1}$ - I $b_{\#2, \#1}$, $a_{\#1, \#2}$]] &, {d, d}]
\end{lstlisting}

In the above definition slots \lstinline/a_/ and \lstinline/b_/ are used to
specify the symbols used to denote the real and the imaginary parts of the
matrix elements. 

Additionally one has to take into account the fact that symbols
\lstinline/a_{i,j}/ and \lstinline/b_{i,j}/ represent real numbers. This fact is
useful during the simplifications in the formulas and can be expressed using the
function
\begin{lstlisting}
SymbolicDensityMatrixAssume[a_, b_, d_] := 
  $\$ $Assumptions = Map[Element[#, Reals] &, 
    Flatten[Join[
        Table[$a_{i,j}$, {i,1,d}, {j,i,d}],
        Table[$b_{i,j}$, {i,1,d}, {j, i+1, d}]
      ]]
  ]
\end{lstlisting}

It is easy to see that the normalization condition can be easily added to the
list of assumptions. However, the conditions for the positivity, \eg\ in the
form of the positivity conditions for the principal minors, are more
complicated~\cite[Chapter~1]{bhatia07positive}.

One should note that, in order to utilize the hermicity conditions for a matrix
defined using function \lstinline/SymbolicDensityMatrix/, is it necessary to
execute function specifying assumptions -- \lstinline/SymbolicDensityMatrixAssume/ -- with the same symbolic
arguments.

Another function useful for the purpose of analyzing the operation on quantum
states is \lstinline/SymbolicMatrix/ function defined as
\begin{lstlisting}
SymbolicMatrix[a_, d1_, d2_] := 
	Array[Subscript[a, #1, #2] &, {d1, d2}]
\end{lstlisting}

Using \lstinline/Flatten/ function in combination with \lstinline/Map/ we can
impose a list of assumptions on the elements of the symbolic matrix. For
example, if one needs to ensure that the elements of the matrix \lstinline/mA/ are real, this
can be achieved as
\begin{lstlisting}
mA = SymbolicMatrix[a, 2, 2];
$\$ $Assumptions = Map[Element[#, Reals] &, Flatten[mA]]
\end{lstlisting} 

\section{Using real density matrices}

Clearly, the representation of the density used in
Section~\ref{sec:preliminaries} is redundant as the off-diagonal element
$a_{i,j} + \mathrm{i} b_{i,j}$ is conjugate to $a_{j,i} - \mathrm{i} b_{j,i}$.
Using this observation we can represent any density matrix as a real matrix with
elements defined as
\begin{equation}
\ComplexToReal[\rho]_{ij} = \left\{
\begin{array}{cc}
\mathbf{Re}\rho_{ij} & i \leq j\\
-\mathbf{Im}\rho_{ij} & i > j\\
\end{array}
\right..
\end{equation}

The above definition can be translated into \MMA\ code as
\begin{lstlisting}
ComplexToReal[denMtx_] := Block[{d = Dimensions[denMtx][[1]]},
  	Array[If[#1 <= #2, Re[denMtx[[#1, #2]]], 
	  -Im[denMtx[[#1, #2]]]] &, {d, d}]
	  ]
\end{lstlisting}

Thus, for a given density matrix, describing $d$-dimensional system we get a
matrix with $n^2$ real elements, instead of a matrix with $n^2$ complex (or
$n^4$ real) elements. Note, that these numbers can be reduced during the
simulation due to the positivity and normalization conditions, but this requires
distinguishing between diagonal and off-diagonal elements.

In the following we denote the map defined by the \lstinline/ComplexToReal/
function as $\ComplexToReal[\cdot]$. One should note that
$\ComplexToReal: \mathbb{M}_n(\Cplx)\mapsto \mathbb{M}_n(\mathbb{R})$.
However, we will only consider multiplication by real numbers as it does not
affect the hermicity of the density matrix.

The real representation of a density matrix contains the same information as the
original matrix. As such it can be used to reconstruct the initial density
matrix. 

Assuming that \lstinline/realMtx/ represents a real matrix obtained as a
representation of the density matrix one can reconstruct the original density
matrix as
\begin{lstlisting}
RealToComplex[realMtx_] := Block[{d = Dimensions[realMtx][[1]]},
  Array[If[#1 < #2, realMtx[[#1, #2]] + I realMtx[[#2, #1]], 
     If[#1 > #2, realMtx[[#2, #1]] - I realMtx[[#1, #2]], 
      realMtx[[#1, #2]]]] &, {d, d}]
  ]
\end{lstlisting}

The map defined by the function \lstinline/RealToComplex/ will be denoted as
$\RealToComplex[\cdot]$. It is easy to see that for any $\rho$ we have
$\ComplexToReal[\RealToComplex[\rho]]=\rho$.

One can also see that maps $\ComplexToReal$ and $\RealToComplex$ are linear if
one considers the multiplication by real numbers only. Thus it can be represented
as a matrix on the Hilbert-Schmidt space of density matrices. Using this
representation one gets
\begin{equation}
\ComplexToReal[\rho]=\res^{-1} \left( M_{\ComplexToReal}\res (\rho)\right)
\end{equation}
where $\res$ is the operation of reordering elements of the matrix into a vector
\cite{miszczak11singular}.

The introduced representation can be utilized to reduce the amount of memory
required during the simulation. For the purpose of modelling the discrete time
evolution of quantum system, one needs to transform the form of quantum maps
into the real representation. For a map $\Phi$ given as a matrix $M_\Phi$ one
obtains its real representation as
\begin{equation}\label{eqn:map-transform}
M_{\ComplexToReal[\Phi]} = M_{\ComplexToReal} M_{\Phi} M_{\RealToComplex}
\end{equation}
One can see that this allows the reduction of the number of multiplication operations
required to simulate the evolution.

\section{Examples}
Let us now consider some examples utilizing maps $\ComplexToReal$ and
$\RealToComplex$. We will focus on the computation involving symbolic
manipulation of states and operations. Only in the last example we use the
statistical properties of density matrices which have to be calculated
numerically.  

\subsection{One-qubit case}
In the simplest case of two-dimensional quantum system, the symbolic density
matrix can be obtained as
\begin{lstlisting}
SymbolicDensityMatrix[a, b, 2]
\end{lstlisting}
which results in
\begin{equation}
\left(
\begin{array}{cc}
 a_{1,1} & a_{1,2}+i b_{1,2} \\
 a_{1,2}-i b_{1,2} & a_{2,2} \\
\end{array}
\right).
\end{equation}

The list of assumptions required to force \MMA\ to simplify the expressions
involving the above matrix can be obtained as
\begin{lstlisting}
SymbolicDensityMatrixAssume[a, b, 2]
\end{lstlisting}
which results in storing the following list
\begin{lstlisting}
{$a_{1,1} \in$ Reals,$a_{1,2} \in$ Reals, $a_{2,2} \in$ Reals, $b_{1,2} \in$ Reals} 
\end{lstlisting}
in the global variable \lstinline/$\$ $Assumptions/.

In \MMA\ the application of map $\ComplexToReal$ on the above matrix results in
\begin{equation}
\left(
\begin{array}{cc}
 \mathbf{Re}\left(a_{1,1}\right) & \mathbf{Re}\left(a_{1,2}\right)-\mathbf{Im}\left(b_{1,2}\right) \\
 \mathbf{Re}\left(b_{1,2}\right)-\mathbf{Im}\left(a_{1,2}\right) & \mathbf{Re}\left(a_{2,2}\right) \\
\end{array}
\right),
\end{equation}
where \lstinline/Re/ and \lstinline/Im/ are the functions for taking the real
and the imaginary parts of the number. Only after using function
\lstinline/FullSimplify/ one gets the expected form of the output
\begin{equation}
\left(
\begin{array}{cc}
 a_{1,1} & a_{1,2} \\
 b_{1,2} & a_{2,2} \\
\end{array}
\right).
\end{equation}

In the one-qubit case it is also easy to check that map $\ComplexToReal$ is
represented by the matrix
\begin{equation}
M_{\ComplexToReal}^{(2)} = \frac{1}{2}\left(
\begin{smallmatrix}
 2 & 0 & 0 & 0 \\
 0 & 1 & 1 & 0 \\
 0 & -i & i & 0 \\
 0 & 0 & 0 & 2 \\
\end{smallmatrix}
\right).
\end{equation}
The matrix representation of the map $\RealToComplex$ reads
\begin{equation}
M_{\RealToComplex}^{(2)} = (M_{\ComplexToReal}^{(2)})^{-1} = \left(
\begin{smallmatrix}
 1 & 0 & 0 & 0 \\
 0 & 1 & i & 0 \\
 0 & 1 & -i & 0 \\
 0 & 0 & 0 & 1 \\
\end{smallmatrix}
\right).
\end{equation} 

The above consideration can be repeated and in the case of three-dimensional
quantum system the matrix representation of the $\ComplexToReal$ map reads
\begin{equation}
M_{\ComplexToReal}^{(3)}=\frac{1}{2}\left(
\begin{smallmatrix}
 2 & 0 & 0 & 0 & 0 & 0 & 0 & 0 & 0 \\
 0 & 1 & 0 & 1 & 0 & 0 & 0 & 0 & 0 \\
 0 & 0 & 1 & 0 & 0 & 0 & 1 & 0 & 0 \\
 0 & -i & 0 & i & 0 & 0 & 0 & 0 & 0 \\
 0 & 0 & 0 & 0 & 2 & 0 & 0 & 0 & 0 \\
 0 & 0 & 0 & 0 & 0 & 1 & 0 & 1 & 0 \\
 0 & 0 & -i & 0 & 0 & 0 & i & 0 & 0 \\
 0 & 0 & 0 & 0 & 0 & -i & 0 & i & 0 \\
 0 & 0 & 0 & 0 & 0 & 0 & 0 & 0 & 2 \\
\end{smallmatrix}
\right).
\end{equation}

\subsection{One-qubit channels}
The main benefit of the real representation of density matrices is the smaller
number of multiplications required to describe the evolution of the quantum
system.

To illustrate this let us consider a bit-flip channel defined by Kraus operators
\begin{equation}
\left\{
\left(
\begin{array}{cc}
 \sqrt{1-p} & 0 \\
 0 & \sqrt{1-p} \\
\end{array}
\right),
\left(
\begin{array}{cc}
 0 & \sqrt{p} \\
 \sqrt{p} & 0 \\
\end{array}
\right)
\right\},
\end{equation}
or equivalently as a matrix
\begin{equation}
M_{BF}^{(2)} = \left(
\begin{smallmatrix}
 1-p & 0 & 0 & p \\
 0 & 1-p & p & 0 \\
 0 & p & 1-p & 0 \\
 p & 0 & 0 & 1-p \\
\end{smallmatrix}
\right).
\end{equation}

The form of this channel on the real density matrices is given by
\begin{equation}
M_{\ComplexToReal}^{(2)} M_{BF}^{(2)} M_{\RealToComplex}^{(2)} = \left(
\begin{smallmatrix}
 1-p & 0 & 0 & p \\
 0 & 1 & 0 & 0 \\
 0 & 0 & 1-2 p & 0 \\
 p & 0 & 0 & 1-p \\
\end{smallmatrix}
\right).
\end{equation}
This map acts on the real density matrix as
\begin{equation}
\left(
\begin{array}{cc}
 p a_{2,2}-(p-1) a_{1,1} & a_{1,2} \\
 (1-2 p) b_{1,2} & p a_{1,1}-(p-1) a_{2,2} \\
\end{array}
\right).
\end{equation}

One should note that in \MMA\ the direct application of the map $\ComplexToReal$ on the output of the channel, \ie\ $M_{R} M_{BF}\res \rho$, results in
\begin{equation}
\left(
\begin{array}{cc}
 \mathbf{Re}\left(p a_{2,2}-(p-1) a_{1,1}\right) & a_{1,2}+2 \mathbf{Im}(p) b_{1,2} \\
 (1-2 \mathbf{Re}(p)) b_{1,2} & \mathbf{Re}\left(p a_{1,1}-(p-1) a_{2,2}\right) \\
\end{array}
\right).
\end{equation}
In order to get the simplified result one needs to explicitly specify
assumptions \lstinline/p $\in$ Reals/. This is important if one aims at testing the
validity of the symbolic computation, as without this assumptions \MMA\ will not
be able to evaluate the result.

\subsection{Werner states}
As the first example of the quantum states of the composite system let us use
the Werner states defined for two-qubit systems as
\begin{equation}
W(a)=\left(
\begin{smallmatrix}
 \frac{a+1}{4} & 0 & 0 & \frac{a}{2} \\
 0 & \frac{1-a}{4} & 0 & 0 \\
 0 & 0 & \frac{1-a}{4} & 0 \\
 \frac{a}{2} & 0 & 0 & \frac{a+1}{4} \\
\end{smallmatrix}
\right).
\end{equation}

The partial transposition transforms $W(a)$ as
\begin{equation}
W(a)^{T_A} = \left(
\begin{smallmatrix}
 \frac{a+1}{4} & 0 & 0 & 0 \\
 0 & \frac{1-a}{4} & \frac{a}{2} & 0 \\
 0 & \frac{a}{2} & \frac{1-a}{4} & 0 \\
 0 & 0 & 0 & \frac{a+1}{4} \\
\end{smallmatrix}
\right),
\end{equation}
and this matrix has one negative eigenvalue for $a>1/3$, which indicates a
presence of quantum entanglement.

In this case the real representation of quantum states reduces one element from
the $W(a)$ matrix and we get
\begin{equation}
\ComplexToReal[W(a)] = \left(
\begin{smallmatrix}
 \frac{a+1}{4} & 0 & 0 & \frac{a}{2} \\
 0 & \frac{1-a}{4} & 0 & 0 \\
 0 & 0 & \frac{1-a}{4} & 0 \\
 0 & 0 & 0 & \frac{a+1}{4} \\
\end{smallmatrix}
\right).
\end{equation}

This matrix has eigenvalues
\begin{equation}
\left\{\frac{1-a}{4},\frac{1-a}{4},\frac{a+1}{4},\frac{a+1}{4}\right\},
\end{equation}
and we have that the sum of smaller eigenvalues is greater than the larger
eigenvalue for $a>1/3$.

\subsection{Partial transposition}
Another important example related to the composite quantum systems is the case
of partial quantum operations. Such operations arise in the situation when one
needs to distinguish between the evolution of the system and the evolution of
the same system threated as a part of a bigger subsystem.

Let us consider the partial transposition of the two-qubit density matrix
\begin{lstlisting}
$\rho$ = SymbolicDensityMatrix[x, y, 4]
\end{lstlisting}
which is given by
\begin{equation}
\rho^{T_A} = \left(
\begin{array}{cccc}
 x_{1,1} & x_{1,2}+i y_{1,2} & x_{1,3}-i y_{1,3} & x_{2,3}-i y_{2,3} \\
 x_{1,2}-i y_{1,2} & x_{2,2} & x_{1,4}-i y_{1,4} & x_{2,4}-i y_{2,4} \\
 x_{1,3}+i y_{1,3} & x_{1,4}+i y_{1,4} & x_{3,3} & x_{3,4}+i y_{3,4} \\
 x_{2,3}+i y_{2,3} & x_{2,4}+i y_{2,4} & x_{3,4}-i y_{3,4} & x_{4,4} \\
\end{array}
\right).
\end{equation}

One can easily check that in this case
\begin{equation}
\ComplexToReal[\rho^{T_A}] = \left(
\begin{array}{cccc}
 x_{1,1} & x_{1,2} & x_{1,3} & x_{2,3} \\
 y_{1,2} & x_{2,2} & x_{1,4} & x_{2,4} \\
 -y_{1,3} & -y_{1,4} & x_{3,3} & x_{3,4} \\
 -y_{2,3} & -y_{2,4} & y_{3,4} & x_{4,4} \\
\end{array}
\right)
\end{equation}
and 
\begin{equation}
(\ComplexToReal[\rho])^{T_A} = \left(
\begin{array}{cccc}
 x_{1,1} & x_{1,2} & y_{1,3} & y_{2,3} \\
 y_{1,2} & x_{2,2} & y_{1,4} & y_{2,4} \\
 x_{1,3} & x_{1,4} & x_{3,3} & x_{3,4} \\
 x_{2,3} & x_{2,4} & y_{3,4} & x_{4,4} \\
\end{array}
\right),
\end{equation}
and thus
\begin{equation}
\ComplexToReal[\rho^{T_A}] \not = (\ComplexToReal[\rho])^{T_A}.
\end{equation}
For this reason one cannot change the order of operations. However, the
explicit form of the partial transposition on the real density matrices can be
found by representing operation of partial transposition as a matrix
\cite{miszczak11singular},
\begin{lstlisting}
ChannelToMatrix[PartialTranspose[#, {2, 2}, {1}] &, 4]
\end{lstlisting}
and using Eq.~(\ref{eqn:map-transform}).

One should note that this method can be used to obtain an explicit form of any
operation of the form $\Phi\otimes \Id$, where $\Id$ denotes the identity
operation of the subsystem. 

\subsection{Partial trace}
The second important example of a partial operation is the partial trace. This
operation allows obtaining the state of the subsystem. 

For two-qubit density matrix we have
\begin{equation}
\tr_A\rho = \left(
\begin{array}{cc}
 x_{1,1}+x_{3,3} & x_{1,2}+x_{3,4}+i \left(y_{1,2}+y_{3,4}\right) \\
 x_{1,2}+x_{3,4}-i \left(y_{1,2}+y_{3,4}\right) & x_{2,2}+x_{4,4} \\
\end{array}
\right).
\end{equation}

One can verify that the operation of tracing-out the subsystem commutes with the map
$\ComplexToReal$ and in this case we have
\begin{equation}
\RealToComplex[\tr_A\ComplexToReal[\rho]] = \tr_A \rho.
\end{equation}
Thus one can calculate the reduced state of the subsystem using the real
value representation.

\subsection{Random real states}
In this section we focus on the statistical properties of the matrices
representing real quantum states. The main difficulty here is that, in contrast
to the random density matrices, real representations can have complex
eigenvalues.

Random density matrices play an important role in quantum information theory and
they are useful in order to obtain information about the average behavior of
quantum protocols. Unlike the case of pure states, mixed states can be drawn
uniformly using different methods, depending on the used probability
measure~\cite{BZ06, miszczak12generating, miszczak13employing}.

One of the methods is motivated by the physical procedure of tracing-out a
subsystem. In a general case, one can seek a source of randomness in a given
system, by studying the interaction of the $n$-dimensional system in question
with the environment. In such situation the random states to model the behaviour
of the system should be generated by reducing a pure state in $N\times
K$-dimensional space. In what follows we denote the resulting probability
measure by $\mu_{N,K}$. 

Using Wolfram language, the procedure for generating random density matrices
with $\mu_{N,K}$ can be implemented as
\begin{lstlisting}
RandomState[n_, k_] := Block[{gM},
  gM = GinibreMatrix[n, k];
  Chop[#/Tr[#]] &@(gM.ConjugateTranspose[gM])
]
\end{lstlisting}
where function \lstinline/GinibreMatrix/ is defined as
\begin{lstlisting}
GinibreMatrix[n_,k_] := Block[{dist},
	dist = NormalDistribution[0,1];
	RandomReal[dist,{n,k}] + I RandomReal[dist,{n,k}]
]
\end{lstlisting}

\subsection{Spectral properties}

\begin{figure}[ht!]
\begin{center}
	\includegraphics[width=.65\textwidth]{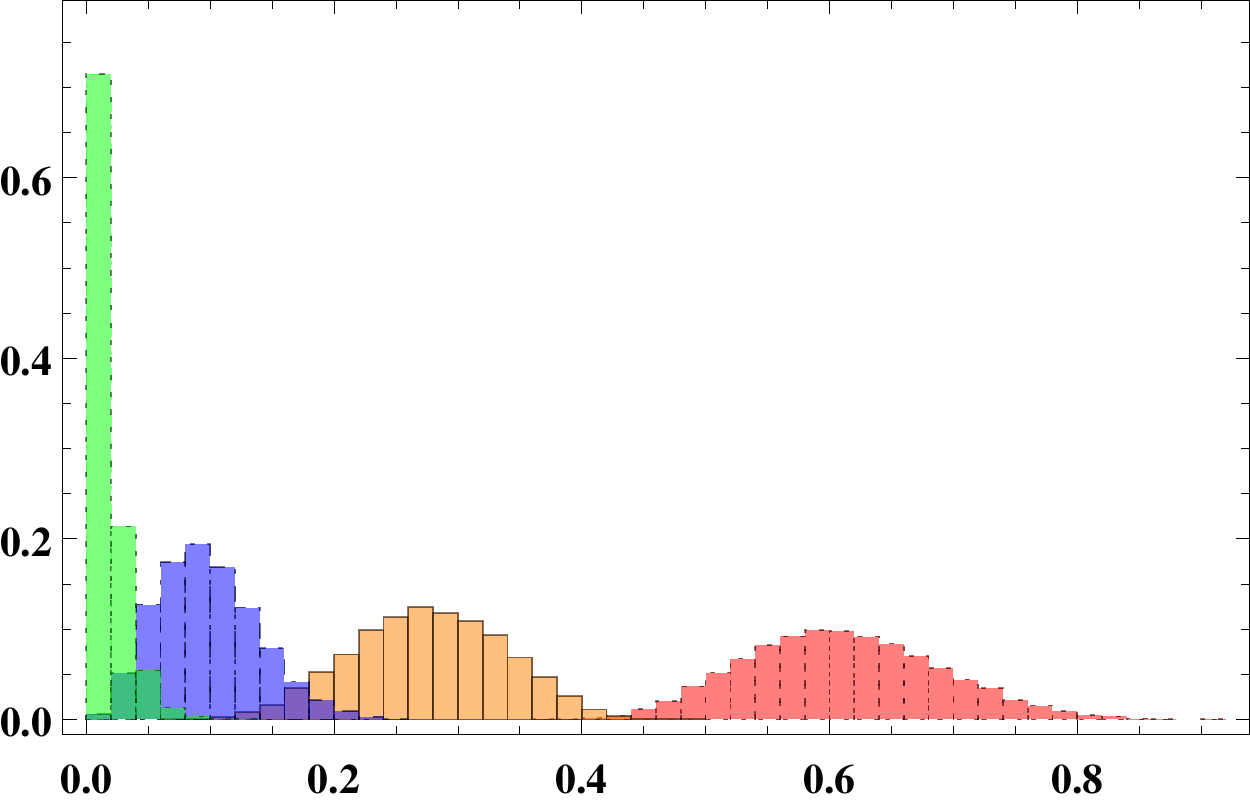}
	\caption{Distribution of eigenvalues for 4-dimensional random density
	matrices distributed uniformly with Hilbert-Schmidt measure for the sample
	of size $10^4$. Each color (and contour style) correspond to the subsequent
	eigenvalue, ordered by their magnitude.}
	\label{fig:histAll}
\end{center}
\end{figure}

In the special case of $K=N$ we obtain the Hilbert-Schmidt ensemble. The
distribution of eigenvalues for $K=N=4$ (i.e. Hilbert-Schmidt ensemble for
ququart) is presented in Fig.~\ref{fig:histAll}.

\begin{figure}[ht!]
\begin{center}
	\subfigure{\includegraphics[width=.4\textwidth]{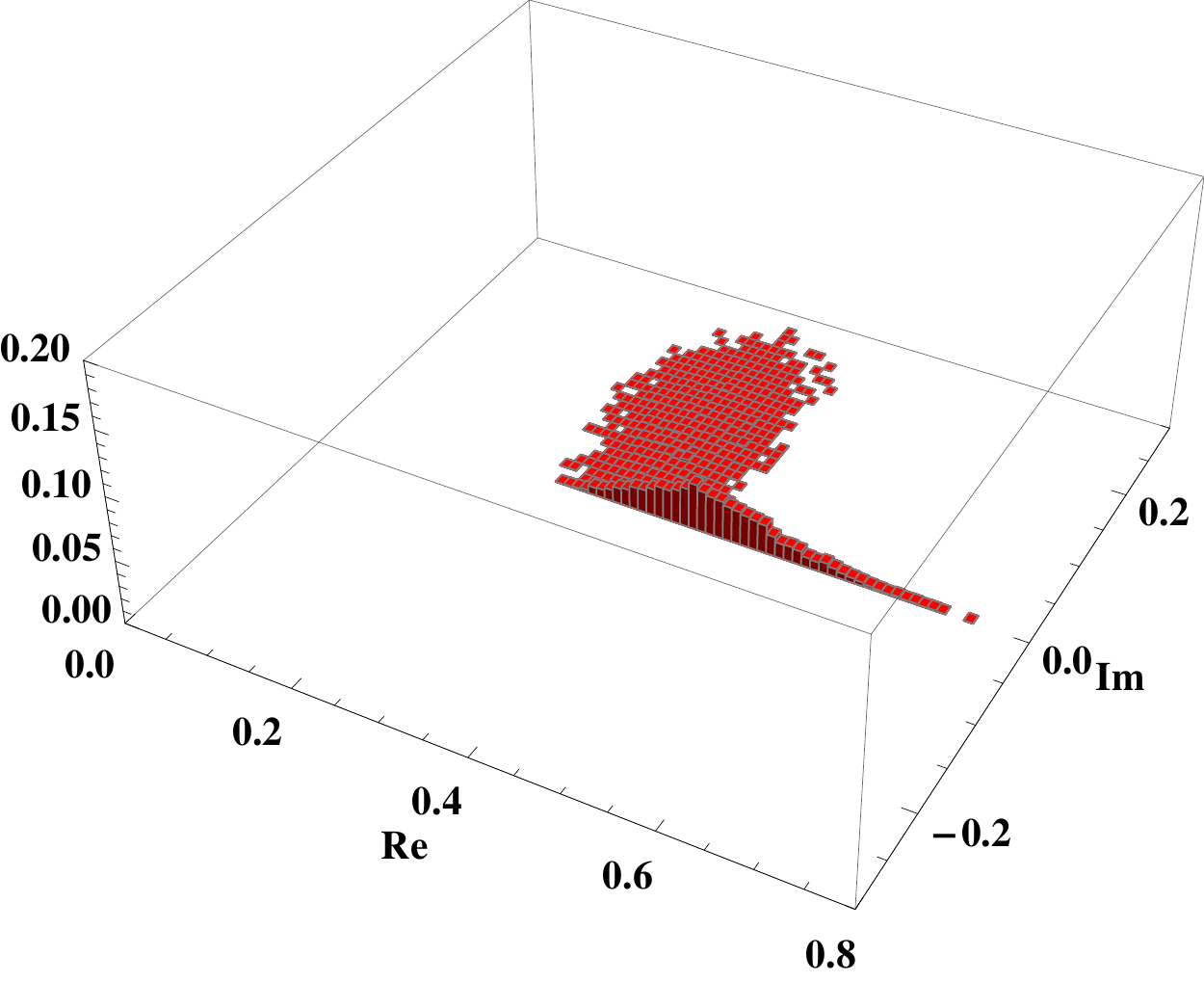}}
	\subfigure{\includegraphics[width=.4\textwidth]{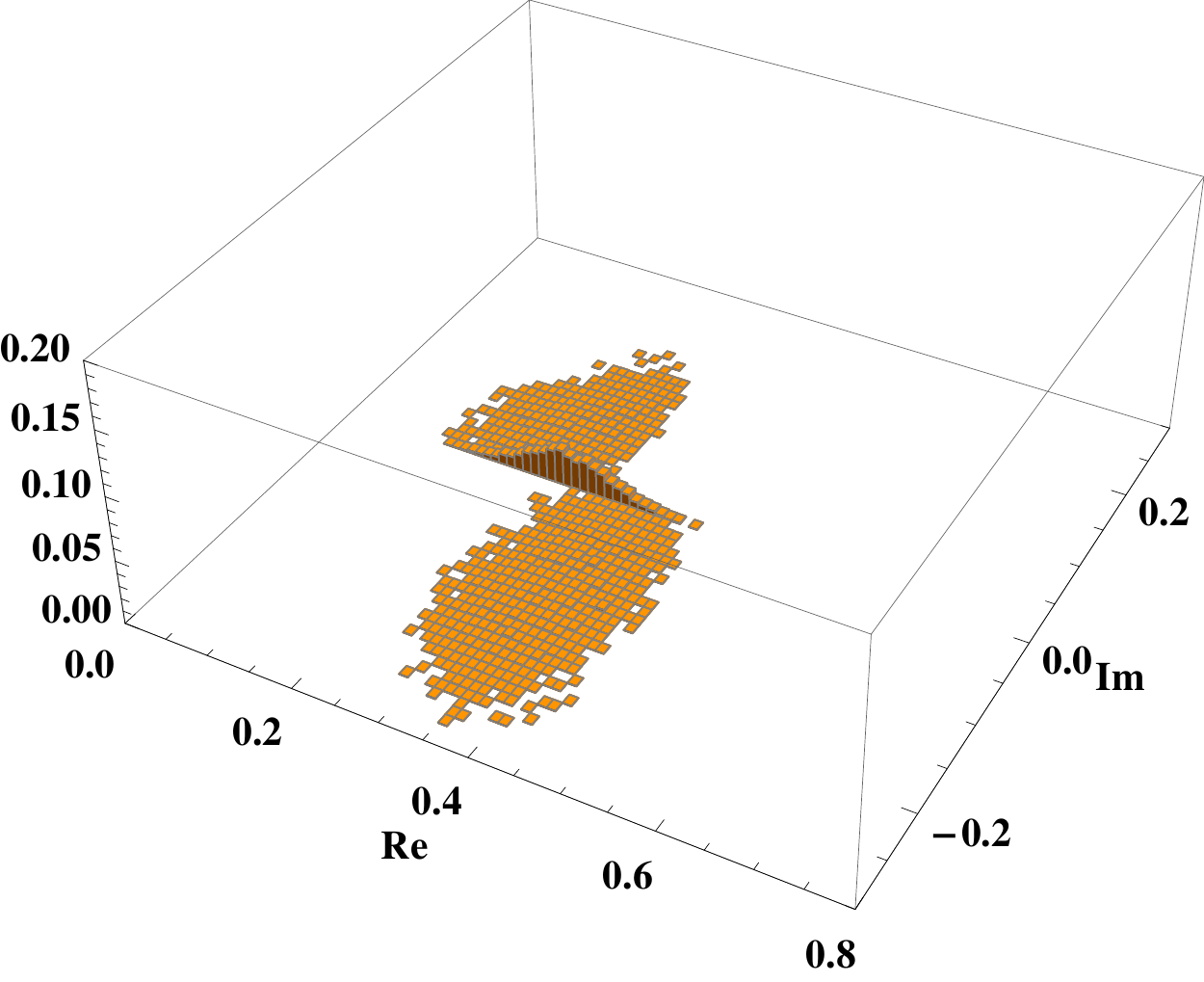}}
	\subfigure{\includegraphics[width=.4\textwidth]{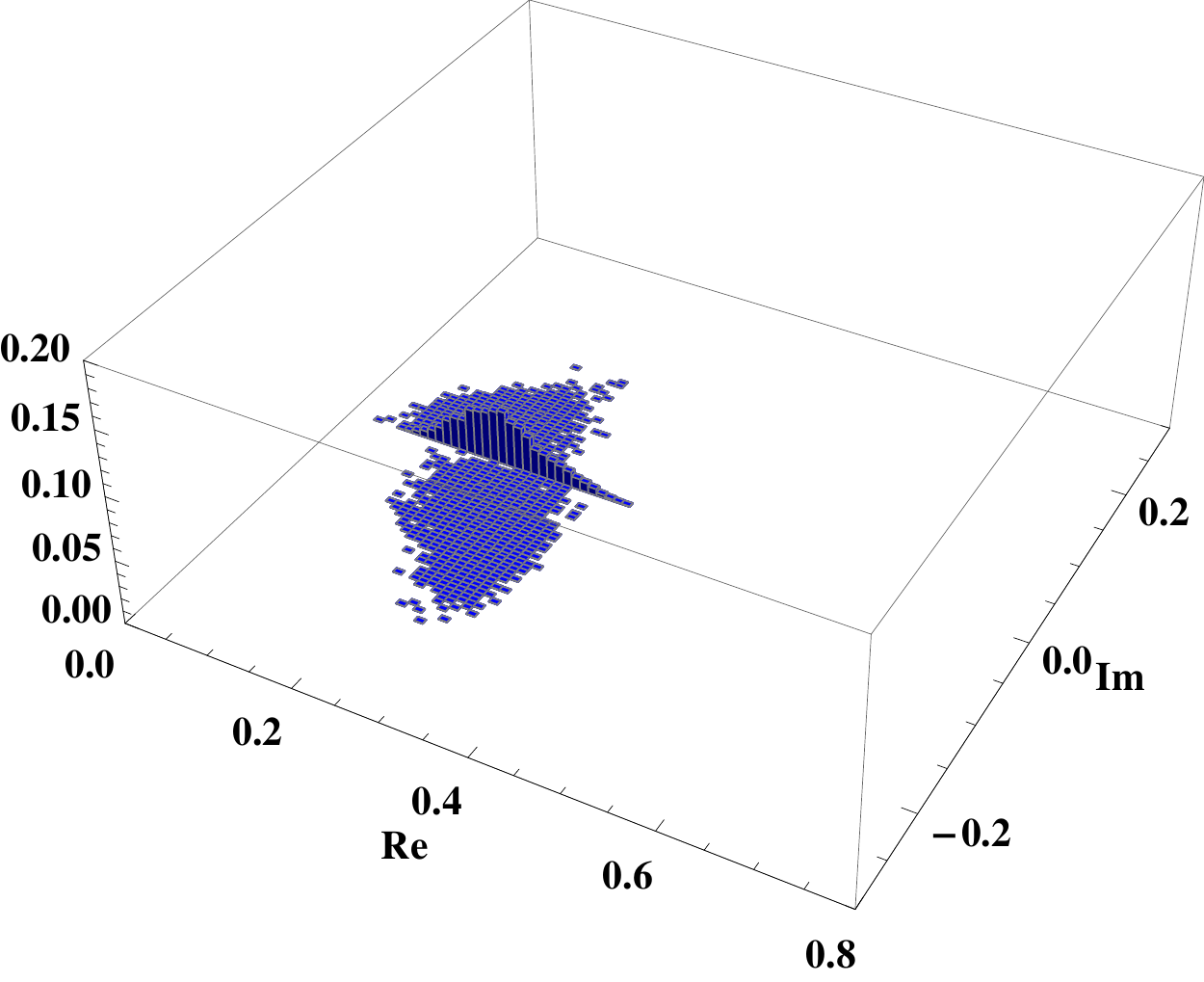}}
	\subfigure{\includegraphics[width=.4\textwidth]{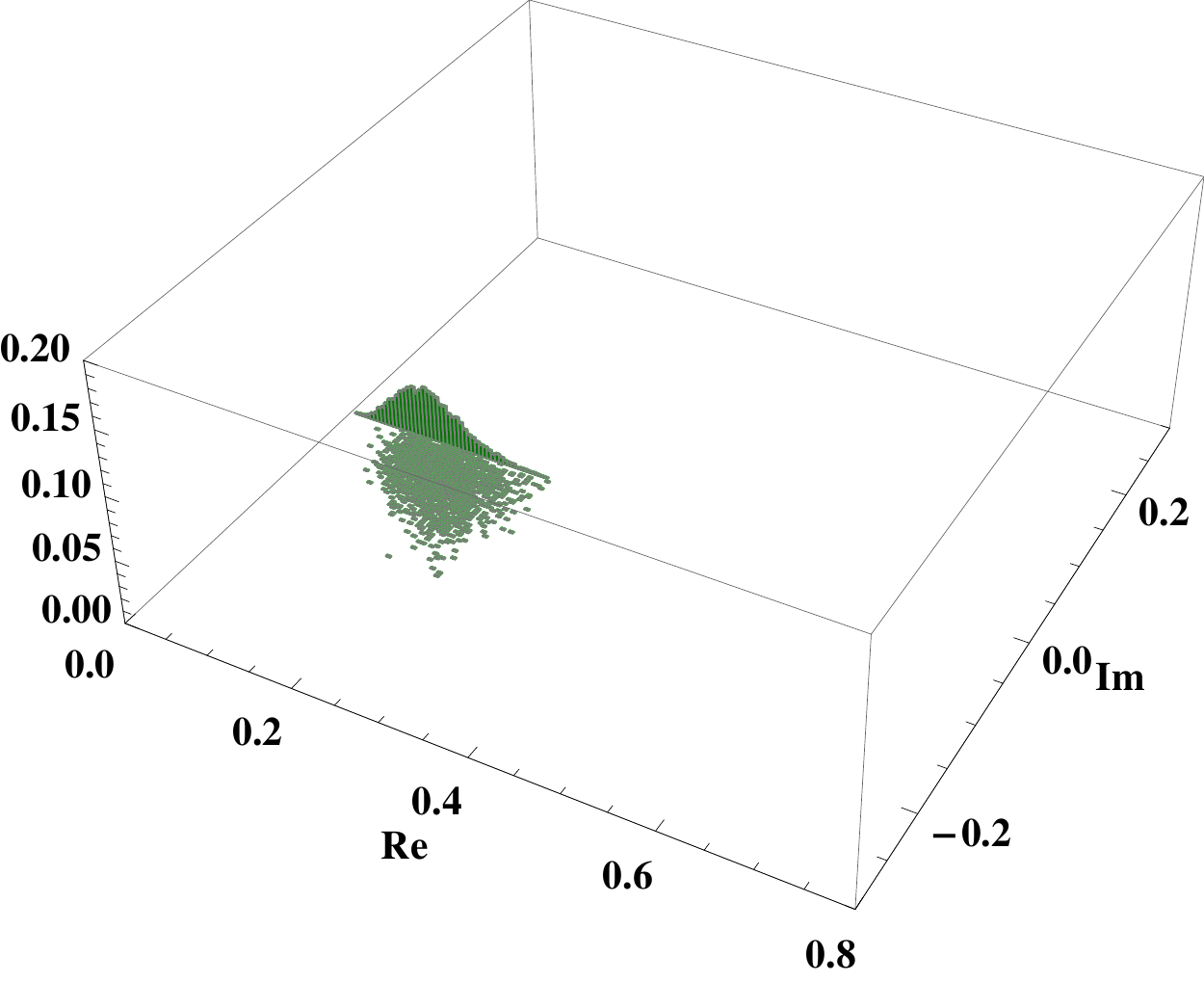}}
	\caption{Distribution of eigenvalues for 4-dimensional random density
	matrices distributed uniformly with Hilbert-Schmidt measure for the sample
	of size $10^4$. Eigenvalues were ordered according to their absolute value.}
	\label{fig:hist3Dx}
\end{center}
\end{figure}

The real representation for the Hilbert-Schmidt ensemble for one ququart
consists of matrices having four eigenvalues. Two of these values are complex
and mutually conjugate (see Fig.~\ref{fig:hist3Dx}). 

\subsubsection{Form of the resulting matrix elements}

Using \lstinline/SymbolicMatrix/ function one can easily analyze the dependency
of the elements of the resulting matrix on the element of the Ginibre matrix.

For the sake of simplicity we demonstrate this on one-qubit states from the
Hilbert-Schmidt ensemble. In this case the Ginibre matrix can be represented as
\begin{lstlisting}
mA = SymbolicMatrix[a, 2, 2];
mB = SymbolicMatrix[b, 2, 2];
m2 = mA + I mB
\end{lstlisting}
The resulting density matrix has (up to the normalization) elements given by the 
matrix 
\begin{lstlisting}
m2.ConjugateTranspose[m2].
\end{lstlisting}
In this case the real representation is given by
\begin{equation}
\left(
\begin{array}{cc}
q_{1,1} & q_{1,2} \\
q_{2,1} & q_{2,2}
\end{array}
\right),
\end{equation}
with
\begin{equation}
\begin{split}
q_{1,1} &= a_{1,1}^2+a_{1,2}^2+b_{1,1}^2+b_{1,2}^2,\\
q_{1,2} &= a_{1,1} a_{2,1}+a_{1,2} a_{2,2}+b_{1,1} b_{2,1}+b_{1,2} b_{2,2},\\
q_{2,1} &= a_{2,1} b_{1,1}+a_{2,2} b_{1,2}-a_{1,1} b_{2,1}-a_{1,2} b_{2,2},\\
q_{2,2} &= a_{2,1}^2+a_{2,2}^2+b_{2,1}^2+b_{2,2}^2.
\end{split}
\end{equation}
Here $a_{i,j}$ and $b_{i,j}$ are independent random variables used in the
definition of the Ginibre matrix.

From the above one can see that the elements of the density matrix resulting
from the procedure for generating random quantum states are obtained as a product
and a sum of the elements of real and imaginary parts of the Ginibre matrix. In
the case of density matrices the normalization imposes the condition $q_{1,1} =
1- q_{2,2}$. Thus, one can also see that the elements are not independent.

\section{Final remarks}
In this note we have introduced a simplified version of quantum states'
representation using the redundancy of information in the standard
representation of density matrices. Our aim was to the find out if such
representation can be beneficial from the point of view of the symbolic
manipulation of quantum states and operations.

To achieve this goal we have used \MMA\ computing system to implement the functions
required to operate on real quantum states and demonstrated some examples where
this representation can be useful from the computational point of view. Its main
advantage is that it can be used to reduce the memory requirements for the
representation of quantum states. Moreover, in some particular cases where the
density matrix contains only real numbers, the real representation reduces to
the upper-triangular matrix. 

The real representation can be also beneficial for the purpose of modelling
quantum channels. Here its main advantage is that it can be used to reduce the
number of multiplications required during the simulation of the discrete quantum
evolution. As a particular example, we have studied the form of partial quantum
operations in the introduced representation. In the case of the partial trace
for the bi-bipartite system, the introduced representation allows the calculation of
the reduced dynamics using the real representation only.

Unfortunately, the introduced representation poses some disadvantages.
The main drawback of the introduced representation is the lack of hermicity of
real density matrices. This makes the analysis of the spectral properties
of real quantum states much more complicated.  

\paragraph{Acknowledgement}
This work has been supported by Polish National Science Centre project number
2011/03/D/ST6/00413 and RAS project on: "Modeling the uncertainty: quantum
theory and imaging processing", LR 7/8/2007. The author would like to thank
G.~Sergioli for motivating discussions.


\end{document}